\documentclass{jetpl}
\usepackage {cite}
\twocolumn

\lat

\title{Stability of the $d$-wave pairing
with respect to the intersite Coulomb repulsion in cuprate
superconductors}

\rtitle{Stability of the $d$-wave pairing with respect to the
intersite  Coulomb repulsion in cuprate superconductors}

\sodtitle{Stability of the $d$-wave pairing with respect to the
intersite Coulomb repulsion in cuprate superconductors}

\author{V.\,V.\,Val'kov$^{a}$,
D.\,M.\,Dzebisashvili$^{a}$, M.\,M.\,Korovushkin$^{a}$, A.\,F.
Barabanov$^{b}$}

\rauthor{V.\,V.\,Val'kov, D.\,M.\,Dzebisashvili,
M.\,M.\,Korovushkin, and A.\,F. Barabanov}

\address{$^{a}$Kirensky Institute of Physics, Federal Research Center KSC SB RAS, 660036 Krasnoyarsk,
Russia\\
$^{b}$Vereshchagin Institute for High Pressure Physics, 108840
Troitsk, Russia}

\abstract{Within the spin-fermion model for cuprate
superconductors, the
 influence of the intersite Coulomb interactions $V_2$ and $V_2'$ between holes
 located at the next-nearest-neighbor oxygen ions of CuO$_2$ plane on the
 implementation of the $d_{x^2-y^2}$-wave pairing is studied. It
 is shown that $d$-wave pairing can be suppressed only for
 unphysically large values of $V_2$ and $V_2'$.
}

\begin{document}

\maketitle

\section{INTRODUCTION}\label{sec1}

\begin{figure}
\begin{center}
\includegraphics[width=0.35\textwidth]{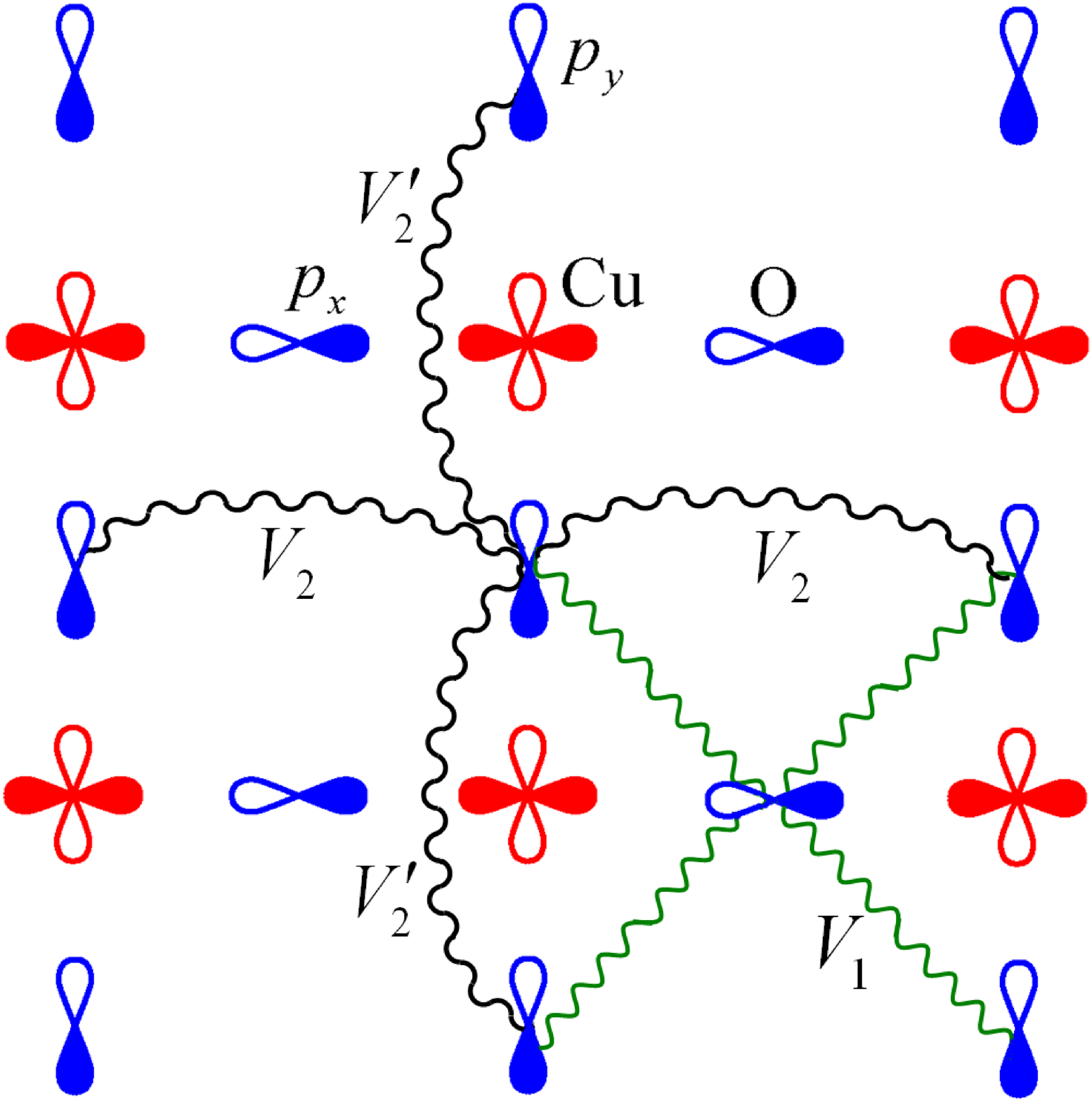}
\caption{Fig. 1. Structure of CuO$_2$ plane. Here $V_1$ denotes
the Coulomb interaction between holes located at the
nearest-neighbor oxygen sites and $V_2$ and $V'_2$ correspond to
the Coulomb interactions of holes located at the
next-nearest-neighbor oxygen sites.} \label{fig-1}
\end{center}
\end{figure}

It is known that the real structure of CuO$_2$ plane is
characterized by the spatial separation of the subsystem of holes
located at oxygen ions and the subsystem of spins localized at
copper ions~(Fig.~\ref{fig-1}). Besides, a number of features is
caused by the presence of two oxygen ions in the unit cell of
copper-oxygen plane. The minimal realistic microscopic model for
cuprates is the three-band $p-d$ model (the Emery
model)~\cite{Emery87,Varma87}. This model takes into account the
$d_{x^2-y^2}$-orbitals of copper ions and $p_x$- and
$p_y$-orbitals of oxygen ions. However, along with the realism,
the multiband character of the Emery model leads to cumbersome
analysis of cuprates physics. That is why a number of studies in
this direction is carried out in the framework of the Hubbard
model and its effective low-energy variants, such as $t-J$ and
$t-J^*$ models on the simple square lattice. In these models, the
same fermions form the charge and the spin subsystems.

Along with the number of important results, such an approach has a
serious disadvantage: the Cooper pairing of fermions caused by the
kinematic~\cite{Zaitsev87}, exchange~\cite{Izyumov9799,Plakida10},
and spin-fluctuation mechanisms considered in the
Hubbard~\cite{Zaitsev04,Val'kov11},
$t-J$~\cite{Izyumov9799,Plakida10}, or
$t-J^*$~\cite{Yushankhai90,Val'kov02} models is suppressed by the
intersite Coulomb repulsion $V_1$ of charge carriers located at
the neighboring sites. This effect is most pronounced in the $d$
channel~\cite{Plakida13} and the Cooper instability disappears
completely at $V_1\sim1-2$ eV.

In our previous paper~\cite{Val'kov16}, it has been shown that,
because of the two-orbital character of the subsystem of holes
located at oxygen sites and the spatial separation of this
subsystem from that of spins at copper ions, the superconducting
phase in high-T$_c$ cuprates is stable with respect to the strong
Coulomb repulsion of holes located at the nearest-neighbor oxygen
sites if the order parameter has the $d_{x^2-y^2}$-symmetry. This
effect is due to the symmetry properties of the Coulomb potential.

Note that in Ref.~\cite{Val'kov16} the stability of the $d$-wave
pairing was proved only for the case of the intersite Coulomb
repulsion of holes located at the nearest-neighbor oxygen
ions,$V_1$, while the role of the Coulomb repulsion between holes
located at the more distant oxygen ions, $V_2$, is still unclear
(the possibility of influence of $V_2$ on the superconducting
$d$-wave pairing has been also mentioned in
Ref.~\cite{Plakida16}). In this paper, we study the role of the
Coulomb interaction between holes located at the
next-nearest-neighbor oxygen ions on CuO$_2$-plane in the
implementation of the superconducting $d_{x^2-y^2}$-wave pairing.

\section{MODEL}\label{sec2}

In the strongly correlated regime, when the Hubbard repulsion
energy $U_d$ is large, i.e., $U_d>\Delta_{pd}\gg t_{pd}$, the
$p-d$ model is reduced to the spin-fermion
model~\cite{Barabanov88,Zaanen88} describing the subsystem of
oxygen holes interacting with the spins located at copper ions.
The Hamiltonian of the spin-fermion model is represented in the
form
\begin{eqnarray}\label{Hamiltonianspinfermion}
&&\hat{H}=\hat{H}_0+\hat{J}+\hat{I}+\hat{V},\\
&&\hat{H}_0=\sum_{k\alpha}\Bigl(\xi_0(k_x)
a_{k\alpha}^{\dagger}a_{k\alpha}+ \xi_0(k_y)
b_{k\alpha}^{\dagger}b_{k\alpha}\nonumber\\
&&\qquad+t_{k}( a_{k\alpha}^{\dagger}b_{k\alpha}+
b_{k\alpha}^{\dagger}a_{k\alpha})\Bigr),\nonumber\\
&&\hat{J}=\frac{J}{N}\sum_{fkq\alpha\beta}e^{if(q-k)}
u_{k\alpha}^{\dag}(\textbf{S}_f\boldsymbol{\sigma_{\alpha\beta}})
u_{q\beta},~~\hat{I}=\frac{I}{2}\sum_{\langle fm\rangle}\textbf{S}_f\textbf{S}_m,\nonumber\\
&&\hat{V}=V_2\sum_{f}\hat{n}_{f+\frac{x}{2}}\hat{n}_{f+\frac{x}{2}+y}+
V_2\sum_{f}\hat{n}_{f+\frac{y}{2}}\hat{n}_{f+\frac{y}{2}+x}\nonumber\\
&&\qquad
+V'_2\sum_{f}\hat{n}_{f+\frac{x}{2}}\hat{n}_{f+\frac{x}{2}+x}+
V'_2\sum_{f}\hat{n}_{f+\frac{y}{2}}\hat{n}_{f+\frac{y}{2}+y},
\end{eqnarray}
where
\begin{eqnarray}\label{Hamiltonian_notations}
&&\xi_0(k_{x(y)})=\varepsilon_p-\mu+\tau(1-\cos
k_{x(y)}),\nonumber\\
&&t_{k}=(2\tau-4t)\sin\frac{k_x}{2}\sin\frac{k_y}{2},\nonumber\\
&&u_{k\beta}=\sin\frac{k_x}{2}a_{k\beta}+
\sin\frac{k_y}{2}b_{k\beta},\nonumber\\
&&\tau=\frac{t_{pd}^2}{\Delta_{pd}}
\biggl(1-\frac{\Delta_{pd}}{U_d-\Delta_{pd}-2V_{pd}}\biggr),\nonumber\\
&&J=\frac{4t_{pd}^2}{\Delta_{pd}}
\biggl(1+\frac{\Delta_{pd}}{U_d-\Delta_{pd}-2V_{pd}}\biggr),\\
&&I=\frac{4t_{pd}^4}{(\Delta_{pd}+V_{pd})^2}\biggl(\frac{1}{U_d}+\frac{2}{2\Delta_{pd}+U_p}\biggr).\nonumber
\end{eqnarray}
The Hamiltonian $\hat{H}_0$ describes the oxygen holes in the
momentum representation. Here $a_{k\alpha}^{\dagger}(a_{k\alpha})$
are the hole creation (annihilation) operators in the oxygen
subsystem with the $p_x$-orbitals~(Fig.~\ref{fig-1}),
$\alpha=\pm1/2$ is the spin projection. Similarly,
$b_{k\alpha}^{\dagger}(b_{k\alpha})$ are operators in the oxygen
subsystem with the $p_y$-orbitals. The bare one-site energy of
oxygen holes is $\varepsilon_p$, $\mu$ is the chemical potential,
and $t$ is the hopping integral. The operator $\hat{J}$ describes
the exchange interaction between the subsystem of oxygen holes and
the subsystem of the spins localized at copper ions. Here,
$\textbf{S}_f$ is the operator of a spin localized at site with
index $f$ and
$\boldsymbol{\sigma}=(\sigma^x,\,\sigma^y,\,\sigma^z)$ is the
vector of the Pauli matrices. The operator $\hat{I}$ describes the
superexchange interaction between the neighboring spins at copper
ions. The intersite Coulomb interaction between holes is described
by the operator $\hat{V}$. As far as the role of the Coulomb
repulsion $V_1$ between holes located at the nearest oxygen sites
was clarified in Ref.~\cite{Val'kov16}, here we do not take into
account the corresponding term in the Hamiltonian $\hat{V}$ and
restrict ourselves to a treatment of the interactions $V_2$ and
$V_2'$ between the next-nearest neighbors (Fig.~\ref{fig-1}). In
the Hamiltonian,
$\hat{n}_{f+x(y)/2}=\sum_{\sigma}\hat{n}_{f+x(y)/2,\sigma}$ is the
operator of the number of holes at the oxygen site $f+x(y)/2$,
where $x=(1,0)$ and $y=(0,1)$ are the lattice basis vectors in the
units of the lattice parameter.

When writing the Hamiltonian (\ref{Hamiltonianspinfermion}), we
take into account that the hopping integrals in the first and the
second terms can have different signs for different hopping
directions owing to the different phases of the wave functions.

Below we use the commonly accepted set of parameters of the model:
$t_{pd}=1.3\,\textrm{eV}$, $\Delta_{pd}=3.6\,\textrm{eV}$,
$U_d=10.5\,\textrm{eV}$,
$V_{pd}=1.2\,\textrm{eV}$~\cite{Hybertsen89,Fischer11,Ogata08}.
Note that for this set, the parameter of the superexchange energy
$I~=~0.136\,\textrm{eV}\,(1570\,\textrm{K})$ agrees well with the
experimental data on cuprate superconductors~\cite{Ogata08}. For
the hopping integral of the holes, we use the value
$t=0.1\,\textrm{eV}$~\cite{Dzebisashvili13}, and we suppose that
the parameters of the intersite Coulomb interactions are
$V_2=V'_2=0.5-1.5\,\textrm{eV}$.

It is important that the exchange energy between the localized and
itinerant spins calculated using the expression
(\ref{Hamiltonian_notations}) is large, namely,
$J=3.4\,\textrm{eV}\gg \tau \approx0.1\,\textrm{eV}$. Therefore,
to describe the oxygen holes dynamics it is necessary to take into
account the exchange interaction rigorously. This problem is
solved using the following basis set of
operators~\cite{Dzebisashvili13,Val'kov14}
\begin{equation}\label{spin-fermion_basis_normal}
a_{k\alpha},\quad b_{k\alpha},\quad
L_{k\alpha}=\frac1N\sum_{fq\beta} e^{if(q-k)}
(\textbf{S}_f\boldsymbol{\sigma}_{\alpha\beta})u_{q\beta},
\end{equation}
where the third operator describes the strong spin-charge
coupling.

\section{EQUATIONS FOR GREEN'S FUNCTIONS}\label{sec3}

For consideration of the conditions for the Cooper instability, we
supplement the basis set (\ref{spin-fermion_basis_normal}) by the
operators ($\bar{\alpha}=-\alpha$)
\begin{eqnarray}\label{spin-fermion_basis_supercond}
a_{-k\bar{\alpha}}^{\dag},\quad b_{-k\bar{\alpha}}^{\dag},\quad
L_{-k\bar{\alpha}}^{\dag}.
\end{eqnarray}
The equations for the normal $G_{ij}$ and the anomalous $F_{ij}$
Green's functions obtained by the method \cite{Zwanzig61,Mori65}
can be represented in the form ($j=1,2,3$)
\begin{eqnarray}\label{equations}
&&(\omega-\xi_{x})G_{1j} = \delta_{1j} + t_{k}G_{2j}+J_{x}G_{3j}
+\Delta_{1k}F_{1j},\nonumber\\
&&(\omega-\xi_{y})G_{2j} = \delta_{2j}+t_{k}G_{1j}+J_{y}G_{3j}
+\Delta_{2k}F_{2j},\nonumber\\
&&(\omega-\xi_{3})G_{3j} = \delta_{3j}K_{k}+
(J_{x}G_{1j}+J_{y}G_{2j})K_{k}
+\Delta_{3k}F_{3j},\nonumber\\
&&(\omega+\xi_{x})F_{1j} = \Delta_{1k}^*G_{1j}
-t_{k}F_{2j}-J_{x}F_{3j},\nonumber\\
&&(\omega+\xi_{y})F_{2j} = \Delta_{2k}^*G_{2j}-
t_{k}F_{1j}-J_{y}F_{3j},\nonumber\\
&&(\omega+\xi_L)F_{3j} =
\Delta^*_{3k}G_{3j}-(J_{x}F_{1j}+J_{y}F_{2j})K_{k}.
\end{eqnarray}
Here, $G_{11}=\langle\langle
a_{k\uparrow}|a_{k\uparrow}^{\dag}\rangle\rangle,\quad
G_{21}=\langle\langle
b_{k\uparrow}|a_{k\uparrow}^{\dag}\rangle\rangle$, and
$G_{31}=\langle\langle
L_{k\uparrow}|a_{k\uparrow}^{\dag}\rangle\rangle$. The functions
$G_{i2}$ and $G_{i3}$ are determined in a similar way with the
only difference that $a^{\dag}_{k\uparrow}$ is replaced by
$b^{\dag}_{k\uparrow}$ and $L^{\dag}_{k\uparrow}$, respectively.
The anomalous Green's functions are defined as
$F_{11}=\langle\langle
a_{-k\downarrow}^{\dag}|a_{k\uparrow}^{\dag}\rangle\rangle,\,
F_{21}=\langle\langle
b_{-k\downarrow}^{\dag}|a_{k\uparrow}^{\dag}\rangle\rangle,\,
F_{31}=\langle\langle
L_{-k\downarrow}^{\dag}|a_{k\uparrow}^{\dag}\rangle\rangle$. For
$F_{i2}$ and $F_{i3}$, the same type of notation regarding the
second index is used. The functions involved in (\ref{equations})
are determined by the expressions
\begin{eqnarray}\label{notations}
&&\xi_{x(y)}=\xi_0({k_{x(y)}}),~~
J_{x(y)}=J\sin\frac{k_{x(y)}}{2},~~K_{k}=\frac{3}{4}-C_1\gamma_{1k},\nonumber\\
&&\xi_L(k)=\varepsilon_p-\mu-2t+5\tau/2-J\nonumber\\
&&\quad+[(\tau-2t)(-C_1\gamma_{1k}+C_2\gamma_{2k})+\tau(-C_1\gamma_{1k}+C_3\gamma_{3k})/2\nonumber\\
&&\quad+JC_1(1+4\gamma_{1k})/4-IC_1(\gamma_{1k}+4)]K_{k}^{-1}.
\end{eqnarray}
Here, $\gamma_{jk}$ are the square lattice invariants:
$\gamma_{1k}=(\cos k_x+\cos k_y)/2,\quad \gamma_{2k}=\cos
k_x\,\cos k_y,\quad\gamma_{3k}=(\cos 2k_x+\cos 2k_y)/2$. In the
course of deriving (\ref{equations}), we assume that the state of
localized moments corresponds to the quantum spin liquid.
Therefore, the spin correlation functions $C_j=\langle
\textbf{S}_0\textbf{S}_{r_j}\rangle$ satisfy the relations
\begin{equation}\label{spin_correlators}
C_j=3\langle S^x_0S^x_{r_j}\rangle=3\langle
S^y_0S^y_{r_j}\rangle=3\langle S^z_0S^z_{r_j}\rangle,
\end{equation}
where $r_j$ is the position of a copper ion within the
coordination sphere $j$. Besides, $\langle S^x_f\rangle=\langle
S^y_f\rangle=\langle S^z_f\rangle=0$.

From (\ref{equations}), it follows that the spectrum of the Fermi
excitations in the normal phase is determined by the solution of
the dispersion equation
\begin{eqnarray}\label{det}
&&\mathrm{det}_{k}(\omega)=(\omega-\xi_{x})(\omega-\xi_{y})
(\omega-\xi_{L})-2J_{x}J_{y}t_{k}K_{k}\nonumber\\
&&-(\omega-\xi_{y})J_{x}^2K_{k} -(\omega-\xi_{x})J_{y}^2K_{k}
-(\omega-\xi_{L})t_{k}^2=0.
\end{eqnarray}
The spectrum is characterized by three bands $\epsilon_{1k}$,
$\epsilon_{2k}$ and $\epsilon_{3k}$~\cite{Val'kov15}. The branch
$\epsilon_{1k}$ with the minimum at a point close to ($\pi/2$,
$\pi/2$) of the Brillouin zone arises owing to the strong
spin-fermion coupling. At the low value of the number of holes per
one oxygen ion $n_p$, the dynamics of holes is determined by the
characteristics of the lower band $\epsilon_{1k}$. This band is
separated by an appreciable gap from the upper bands
$\epsilon_{2k}$ and $\epsilon_{3k}$~\cite{Val'kov15}.

The introduced order parameters $\Delta_{j,k}$ are related to the
anomalous averages as follows
\begin{eqnarray}\label{Deltas}
&&\Delta_{1k}=-\frac{2}{N}\sum_{q}\Bigl(V_2\cos(k_y-q_y)\nonumber\\
&&\qquad\qquad\qquad~+V'_2\cos(k_x-q_x)\Bigr)\langle
a_{q\uparrow}a_{-q\downarrow}\rangle, \nonumber\\
&&\Delta_{2k}=-\frac{2}{N}\sum_{q}\Bigl(V_2\cos(k_x-q_x)\nonumber\\
&&\qquad\qquad\qquad~+V'_2\cos(k_y-q_y)\Bigr)\langle
b_{q\uparrow}b_{-q\downarrow}\rangle, \nonumber
\end{eqnarray}
\begin{eqnarray}
\label{Delta3I} &&\Delta_{3k}=\frac
1N\sum_{q}I_{k-q}\biggl\{\langle
L_{q\uparrow}L_{-q\downarrow}\rangle +C_{1x}\langle
a_{q\uparrow}a_{-q\downarrow}\rangle\nonumber\\
&&\quad+C_{1y}\langle
b_{q\uparrow}b_{-q\downarrow}\rangle+C_1\psi_q\bigl(\langle
a_{q\uparrow}b_{-q\downarrow}\rangle+\langle
b_{q\uparrow}a_{-q\downarrow}\rangle\bigr)\biggr\}K_q^{-1}\nonumber\\
&&\quad-\frac1N\sum_q\biggl\{V_2(C_1\cos k_y-C_2\gamma_{2k})\cos
q_y\nonumber\\
&&\quad+V'_2\left(-\frac{3}{8}+C_1\cos k_x-\frac{C_3}{2}\cos
2k_x\right)\cos q_x\biggr\}K_q^{-1}\langle
a_{q\uparrow}a_{-q\downarrow}\rangle\nonumber\\
&&\quad-\frac1N\sum_q\biggl\{V_2(C_1\cos k_x-C_2\gamma_{2k})\cos
q_x\\
&&\quad+V'_2\left(-\frac{3}{8}+C_1\cos k_y-\frac{C_3}{2}\cos
2k_y\right)\cos q_y\biggr\}K_q^{-1}\langle
b_{q\uparrow}b_{-q\downarrow}\rangle.\nonumber
\end{eqnarray}
Here $C_{1x(1y)}=C_1\displaystyle\sin^2\frac{q_{x(y)}}{2}$,
$\psi_{k}=\displaystyle\sin\frac{k_x}{2}\sin\frac{k_y}{2}$ and
$I_k=4I\gamma_{1k}$.

\section{EQUATIONS FOR THE SUPERCONDUCTING ORDER PARAMETERS}\label{sec4}

For the analysis of the conditions for the appearance of the
Cooper instability, we express the anomalous Green's functions in
terms of the $\Delta^*_{lk}$ parameters in the linear
approximation
\begin{eqnarray}\label{Fij}
&&F_{nm}(k,\omega)=\sum_{l=1}^3S^{(l)}_{nm}(k,\omega)\Delta_{lk}^*/\textrm{Det}_k(\omega),
\end{eqnarray}
where
\begin{eqnarray}
&&\textrm{Det}_k(\omega)=-\textrm{det}_k(\omega)\textrm{det}_k(-\omega),\nonumber\\
&&S^{(1)}_{11}(k,\omega)= Q_{3y}(k,\omega)Q_{3y}(k,-\omega),\nonumber\\
&&S^{(2)}_{11}(k,\omega)=S^{(1)}_{22}(k,\omega)= Q_{3}(k,\omega)Q_{3}(k,-\omega),\nonumber\\
&&S^{(1)}_{33}(k,\omega)=K_kS^{(3)}_{11}(k,\omega)=K_k^2Q_{y}(k,\omega)Q_{y}(k,-\omega),\nonumber\\
&&S^{(2)}_{22}(k,\omega)=Q_{3x}(k,\omega)Q_{3x}(k,-\omega),\nonumber\\
&&S^{(2)}_{33}(k,\omega)=K_kS^{(3)}_{22}(k,\omega)=K_k^2Q_{x}(k,\omega)Q_{x}(k,-\omega),\nonumber\\
&&S^{(1)}_{12}(k,\omega)=S^{(1)}_{21}(k,-\omega)=Q_{3}(k,\omega)Q_{3y}(k,-\omega),\nonumber\\
&&S^{(2)}_{12}(k,\omega)=S^{(2)}_{21}(k,-\omega)=Q_{3}(k,\omega)Q_{3x}(k,-\omega),\nonumber\\
&&S^{(3)}_{12}(k,\omega)=S^{(3)}_{21}(k,-\omega)=K_kQ_{x}(k,\omega)Q_{y}(k,-\omega),\nonumber\\
&&S^{(3)}_{33}(k,\omega)=K_kQ_{xy}(k,\omega)Q_{xy}(k,-\omega).
\end{eqnarray}
The functions used here are
\begin{eqnarray}
&&Q_{x(y)}(k,\omega)=(\omega-\xi_{x(y)})J_{y(x)}+t_kJ_{x(y)},\nonumber\\
&&Q_3(k,\omega)=(\omega-\xi_L)t_k+J_xJ_yK_k,\nonumber\\
&&Q_{3x(3y)}(k,\omega)=(\omega-\xi_L)(\omega-\xi_{x(y)})-J_{x(y)}^2K_k,\nonumber\\
&&Q_{xy}(k,\omega)=(\omega-\xi_x)(\omega-\xi_y)-t_k^2.
\end{eqnarray}
Using the spectral theorem~\cite{Zubarev60}, we find the
expressions for the anomalous averages and finally arrive at the
closed set of uniform integral equations for the superconducting
order parameters ($l=1,2,3$)
\begin{eqnarray}\label{Deltas_spectral_theorem}
&&\Delta_{1k}^*=\frac
{2}{N}\sum_{lq}(V_2\cos(k_y-q_y)+V_2'\cos(k_x-q_x))M^{(l)}_{11}(q)
\Delta_{lq}^*,\nonumber\\
&&\Delta_{2k}^*=\frac
{2}{N}\sum_{lq}(V_2\cos(k_x-q_x)+V_2'\cos(k_y-q_y))M^{(l)}_{22}(q)
\Delta_{lq}^*,\nonumber\\
&&\Delta_{3k}^*=\frac{1}{N}\sum_{lq}\left\{I_{k-q}\Bigl[
C_{1x}M^{(l)}_{11}(q)+C_{1y}M^{(l)}_{22}(q)-M^{(l)}_{33}(q)\right.\nonumber\\
&&~+C_1\phi_q\bigl(M^{(l)}_{12}(q)+M^{(l)}_{21}(q)\bigr)\Bigr]\nonumber\\
&&~+\Bigl[V_2(C_1\cos k_y-C_2\gamma_{2k})\cos
q_y\nonumber\\
&&~+V_2'\left(-\frac{3}{8}+C_1\cos k_x-\frac{C_3}{2}\cos
2k_x\right)\cos q_x\biggr]M^{(l)}_{11}(q)\nonumber\\
&&~+\Bigl[V_2(C_1\cos k_x-C_2\gamma_{2k})\cos
q_x\\
&&~\left.+V_2'\left(-\frac{3}{8}+C_1\cos k_y-\frac{C_3}{2}\cos
2k_y\right)\cos
q_y\biggr]M^{(l)}_{22}(q)\right\}\frac{\Delta_{lq}^*}{K_q},\nonumber
\end{eqnarray}
where
$$ M^{(l)}_{nm}(q)=\frac{S^{(l)}_{nm}(q,E_{1q})
+S^{(l)}_{nm}(q,-E_{1q})}{4E_{1q}(E_{1q}^2-E_{2q}^2)
(E_{1q}^2-E_{3q}^2)}\tanh\left(\frac{E_{1q}}{2T}\right).
$$
Below, we use the system (\ref{Deltas_spectral_theorem}) to find
the critical superconducting temperature.
\begin{figure}[t]
\begin{center}
\includegraphics[width=0.47\textwidth]{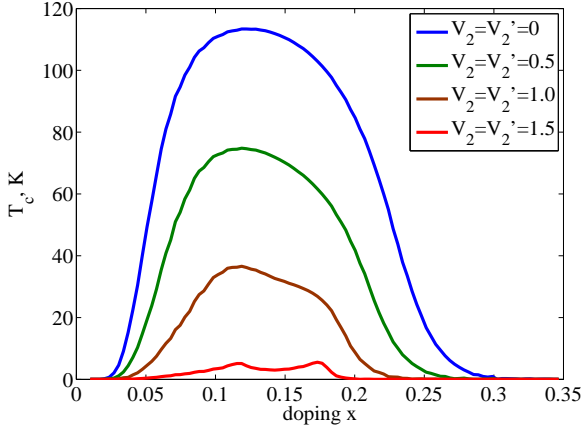}
\caption{Fig. 2. Critical temperature for the transition to the
superconducting $d_{x^2-y^2}$ phase versus doping at four values
of the Coulomb repulsion parameter $V_2$ and $V_2'$.}\label{fig-2}
\end{center}
\end{figure}

In the Fig.~\ref{fig-2}, we illustrate the results obtained by
solving Eq.~\ref{Deltas_spectral_theorem} for the
$d_{x^2-y^2}$-wave pairing, where
\begin{eqnarray*}
\Delta_{lk}=\Delta_{l1}\cdot(\cos k_x-\cos
k_y)+\Delta_{l2}\cdot(\cos 2k_x-\cos 2k_y).
\end{eqnarray*}
One can see from Fig.~\ref{fig-2} that an increase in $V_2$ and
$V_2'$ leads to suppression of the $d$-wave pairing, however
superconductivity is maintained up to unphysically large values
$V_2=V_2'=1.5$ eV of the Coulomb interaction between holes located
at the next-nearest-neighbor oxygen ions (for comparison, the
intensity of the Coulomb interaction between nearest-neighbor
oxygen ions $V_1=1-2$ eV~\cite{Fischer11}).

\section*{CONCLUSION}

To conclude, we have shown that the intersite Coulomb repulsion
between holes located at the next-nearest-neighbor oxygen ions of
CuO$_2$ plane suppresses the $d_{x^2-y^2}$-wave pairing only at
unphysically large values of the Coulomb interaction
$V_2=V_2'=1.5$ eV. Taking into account our previous
result~\cite{Val'kov16} on cancelation of the effect of the
Coulomb interaction $V_1$ for the nearest-neighbor oxygen sites on
the $d$-wave pairing, we conclude that an account for the real
structure of CuO$_2$ plane leads to stability of the
$d_{x^2-y^2}$-wave pairing towards the strong intersite Coulomb
repulsion.  It is obvious that an account for the Coulomb
interaction $V_3$ does not effect on the superconducting $d$-wave
pairing because of the same "symmetry reason" as that for
$V_1$~\cite{Val'kov16}.

The work was supported by the Russian Foundation for Basic
Research (RFBR) and partly by the Government of Krasnoyarsk Region
(project nos. 16-42-240435 and 16-42-243057). The work of A.F.B.
was supported by RFBR (project no. 16-02-00304). The work of
M.M.K. was supported by grant of the President of the Russian
Federation (project MK-1398.2017.2).

\end{document}